\documentclass[preprint,floats,aps,epsfig,nofootinbib,amssymb]{revtex4-1}
\pdfoutput=1
\usepackage{mathrsfs}

\usepackage{slashed}
\usepackage{graphicx,color}
\usepackage{dcolumn}
\usepackage{bm}
\usepackage{subfig}
\usepackage{graphicx}
\usepackage{amssymb}

\begin{document}

\preprint{ACFI-T15-07}

\title{ATLAS Diboson Excesses from the Stealth Doublet Model }

\author{Wei Chao}
\email{chao@physics.umass.edu}

\affiliation{
Amherst Center for Fundamental Interactions, Department of Physics, University of Massachusetts-Amherst
Amherst, MA 01003 
 }

\begin{abstract}

The ATLAS collaboration has reported excesses in diboson invariant mass searches of new resonances around 2 TeV, which might be a prediction of new physics around that mass range. We interpret these results in the context of a modified stealth doublet model where the extra Higgs doublet has a Yukawa interaction with the first generation quarks, and show that the heavy CP-even Higgs boson can naturally explain the excesses in the $WW$ and $ZZ$ channels with a small Yukawa coupling, $ \xi\sim 0.15 $, and a tiny mixing angle with the SM Higgs boson, $\alpha \sim 0.06$. Furthermore, the model satisfy  constraints from colliders and electroweak precision measurements.

\end{abstract}

\draft

\maketitle
\section{introduction}

Excesses in searching for diboson resonance using boson-tagged jets were recently reported by the ATLAS collaboration~\cite{Aad:2015owa}. It shows local excesses in the $WZ$, $WW$ and $ZZ$ channels with significance of $3.4\sigma$, $2.6\sigma$ and $2.9\sigma$ respectively.  Similarly, the CMS collaboration~\cite{Khachatryan:2014hpa,Khachatryan:2014gha} has reported an excess of $1.9\sigma$ significance in the dijet resonance channel and $e\nu \bar b b$ channel which may arise from $Wh$ with $h$ decaying hadronically.  These excesses may be evidences of new symmetries or new particles  near $2~{\rm TeV}$.

Since the resonances decay into two gauge boson, they should be bosonic states. Possible origins of this excess were studied by several groups~\cite{Hisano:2015gna,Fukano:2015hga,Franzosi:2015zra,Cheung:2015nha,Dobrescu:2015qna,Aguilar-Saavedra:2015rna,Gao:2015irw,Thamm:2015csa,Brehmer:2015cia,Cao:2015lia,Cacciapaglia:2015eea,Abe:2015jra,Allanach:2015hba,Abe:2015uaa,Carmona:2015xaa,Dobrescu:2015yba,Chiang:2015lqa,Cacciapaglia:2015nga,Sanz:2015zha,Chen:2015xql}, where  the excesses were explained  as spin-1 gauge bosons~\cite{Hisano:2015gna,Cheung:2015nha,Dobrescu:2015qna,Gao:2015irw,Thamm:2015csa,Brehmer:2015cia,Cao:2015lia,Cacciapaglia:2015eea,Abe:2015jra,Dobrescu:2015yba} in an extended gauge group, composite spin-1 resonances~\cite{Fukano:2015hga,Franzosi:2015zra,Carmona:2015xaa}, spin-0 or spin-2 composite particles~\cite{Chiang:2015lqa,Cacciapaglia:2015nga,Sanz:2015zha} and extra scalar bosons~\cite{Chen:2015xql}.   The key points in explaining the excesses are the interactions of new resonance with the Standard Model (SM) gauge bosons, quarks and(or) gluons, the former of which is relevant to the branching ratio of the new resonance and the latter of which is relevant to the production of the new resonance at the LHC. One the one hand, one needs the couplings of new interactions to be large enough  so as to give rise to a sizable production cross section at the LHC, on the other hand the strengths of these interactions should be consistent with current constraints of colliders and electroweak precision measurements. These two requirements are mutual restraint.  A new resonance is not able to explain the ATLAS excesses if its interaction strengths are not mutually compatible with these two requirements.

In this paper, we explain the ATLAS excesses  in the stealth doublet model, where the second Higgs doublet, $H_2$, gets no vacuum expectation value, with mass near 2 TeV, and only the CP-even part of $H_2$ mixes with the SM Higgs boson.  We assume $H_2$ has sizable Yukawa interaction with the first generation quarks, which is consistent with constraints of flavor physics. Such that the heavy CP-even Higgs boson can be produced at the LHC via the Yukawa interaction and decays into diboson states through the mixing with the SM Higgs boson.  Our numerical simulations show that one has $\sigma(pp\to H\to WW/ZZ)\sim 5~{\rm fb}$ by setting $\xi\sim0.15$ and $\alpha\sim0.06$, where $\xi $ is the Yukawa coupling of the $H_2$ with the first generation quarks and $\alpha$ is the mixing angle between two CP-even neutral states. This result is consistent with current constraints from colliders and electroweak precision measurements. 

The remaining of the paper is organized as follows: In section II we give a brief introduction to the model. Section III is the study of constraints on the model. We investigate the ATLAS diboson excesses arising from this stealth doublet model in section IV.  The last part is the concluding remarks.

\section{the model}
We work in the modified stealth doublet model~\cite{Enberg:2013jba,Enberg:2013ara}, where the second Higgs doublet gets no vacuum expectation value (VEV) but its CP-even part mixes with the SM Higgs boson. In the following, we describe the modified stealth doublet model first, and then study its implications in the ATLAS diboson excesses.  The Higgs potential is the same as that in the general two Higgs doublet model (2HDM), which can be written as
\begin{eqnarray}
V&=& -m_1^2 H_1^\dagger H_1^{} + m_2^{} H_2^\dagger H_2^{}  +\left(  {m_{12}^2 H_1^\dagger H_2^{} }+ {\rm h.c.} \right)
\nonumber \\&&+ \lambda_1 (H_1^\dagger H_1)^2 + \lambda_2  (H_2^\dagger H_2^{} )^2 + \lambda_3 (H_1^\dagger H_1^{} ) (H_2^\dagger H_2^{}) + \lambda_4 (H_1^\dagger H_2^{} )(H_2^\dagger H_1^{}) \nonumber \\
&&+ \left\{ {1\over 2 }\lambda_5 (H_1^\dagger H_2^{} )^2 + (\lambda_6 H_1^\dagger H_1+ \lambda_7 H_2^\dagger H_2 ) H_1^\dagger H_2+ {\rm h.c.}   \right\} \label{potential}
\end{eqnarray}
In this paper, we assume the Higgs potential is CP-conserving, so all couplings in eq.(\ref{potential}) are real.   Only one Higgs doublet gets nonzero VEV in the stealth doublet model, we take it be $H_1$. The tadpole conditions for the electroweak symmetry breaking become
\begin{eqnarray}
m_1^2 = \lambda_1^{} v_1^2 \;  ,\hspace{1cm}  m_{12}^2 = -{1\over 2 } \lambda_6^{} v_1^2 
\end{eqnarray} 
where $v_1=\sqrt{2}\langle H_1^{} \rangle \approx 246~{\rm GeV}$.  After spontaneous breaking of the electroweak symmetry, there are two CP-even scalars $h$ and $H$, one CP-odd scalar $A$ and two charged scalars $C^\pm$, the mass eigenvalues of which can be written as~\cite{Enberg:2013jba}
\begin{eqnarray}
m_{A~~}^2 & =& m_2^2 +{1\over 2 } (\lambda_3+\lambda_4-\lambda_5) v_1^2   \\
m_{C~~}^2 &=& m_2^2 + {1\over 2 } \lambda_3 v_1^2  \\
m_{h,H}^2 &=&{1\over 2 }\left\{ m_1^{2} + m_A^2 +\lambda_5 v_1^2 \pm \sqrt{ (m_1^2 -m_A^2 -\lambda_5v_1^2 )^2 -4 \lambda_6^2 v_1^4}\right\}
\end{eqnarray}
The mixing angle $\alpha$ between $h$ and $H$ can be calculated directly, we take it as a new degree of freedom in this paper. $H$ interacts with dibosons through the mixing. We refer the reader to Ref. \cite{Enberg:2013jba} for the feynman rules of Higgs interactions. 

The Yukawa interactions of $H_1$ with SM fermions are exactly the same as  Yukawa interactions of the SM Higgs with fermions in the SM. We assume $H_2$  has sizable Yukawa coupling with the first generation quarks:
\begin{eqnarray}
{\cal L}_N = \sqrt{2} \xi \overline{Q_1} \tilde H_2 u_R^{} + {\rm h.c.} \; .  \label{yukawah}
\end{eqnarray}
where $Q_1 =(u_L, ~d_L)^T$ and $\tilde H_2 =i \sigma_2 H_2^*$.
Since $\langle H_2 \rangle =0$, there is almost no constraint on this Yukawa coupling, and $H$ can be produced at the LHC via this interaction.

\section{Constraints}
Before proceeding to study ATLAS diboson excesses, let us investigate  constraints on the mixing angle $\alpha$.  Couplings of the SM-like Higgs to  other SM particles were measured by the ATLAS and CMS collaboration. Comparing with SM Higgs couplings, couplings of $h$ and $ H$ to all SM states (except $u$ quark) are rescaled by $\cos \alpha$ and $\sin \alpha$, respectively:
\begin{eqnarray}
g_{hXX} = \cos\alpha g_{hXX}^{SM}\; , \hspace{1cm} g_{HXX}=\sin\alpha g_{hXX}^{\rm SM}
\end{eqnarray} 
where $X$ represents SM states. 
Thus signal rates of the Higgs measurements relative to SM Higgs expectations are the function of $\cos \alpha$. Performing a global $\chi^2 $ fit to the Higgs data given by ATLAS and CMS, one has $\cos\alpha \geq 0.84$~\cite{Profumo:2014opa}, at the $95\%$ confidence level.

Another constraint comes from the oblique parameters~\cite{Peskin:1990zt,Peskin:1991sw}, which are defined in terms of contributions to the vacuum polarizations of gauge bosons. The explicit expressions of $\Delta S$ and $\Delta T$, which involve effects of all scalars, can be written as~\cite{Haber:2010bw}
\begin{eqnarray}
\Delta S&=&{1\over \pi m_Z^2 } \left\{ (-1)^i s^2\sum_{i} [ {\cal B }_{2} (m_Z^2; m_Z^2, m_i^2 ) -m_Z^2 {\cal B}_0 (m_Z^2; m_Z^2, m_i^2 )  ]  + c^2 {\cal B}_{2} (m_Z^2; m_H^2, m_{A}^{2} )\right. \nonumber \\ \label{scal}
&&\left.+ s^2 {\cal B}_{2} (m_Z^2; m_h^2, m_{A}^{ 2} )-{\cal B}_{2} (m_Z^2; m_C^2,m_C^2){\over}\right\}\\
\Delta T&=& {1\over 4\pi s_W^2 m_W^2 }\{ s^2 B_{2} (0; m_C^2, m_h^2)+c^2 B_{2}(0;m_C^2,m_H^2) + B_{2} (0;m_C^2,m_{A}^{2}) -s^2 B_{2} (0; m_h^2, m_{A}^2 )\nonumber \\
&&-c^2 B_{2}(0;m_H^2,m_A^2) -s^2 B_{2} (0;m_W^2, m_h^2 ) +s^2 B_{2} (0; m_W^2, m_H^2 )+s^2 B_{2} (0; m_Z^2,m_h^2 )\nonumber \\&& -s^2 B_{2}(0; m_Z^2, m_H^2)+m_W^2 s^2 [B_0(0;m_W^2,m_h^2 )-B_0 (0; m_W^2, m_H^2 )]\nonumber \\&&+M_Z^2  s^2 [- B_0 (0,m_W^2, m_h^2 ) +  B_0(0;m_Z^2,m_H^2)]-{1\over 2 } A_0(m_C^2 ) \} \label{tcal}
\end{eqnarray}
where ${\cal B}_{i} (x; y, z) = B_{i} (x; y,z)-B_{i} (0;y,z)$, $i=(0,~2)$, the expressions of $B_i(x;y,z)$ and $A_0(x)$ can be find in Ref.~\cite{Haber:2010bw}, $c=\cos\alpha$ and $s=\sin \alpha$, $s_W=\sin \theta_W$ with $\theta_W$ the weak mixing angle, $M_Z$ and $M_W$ are masses of $Z$ and $W$ bosons respectively.

\begin{figure}[t!]
\centering
\includegraphics[width=0.46\textwidth]{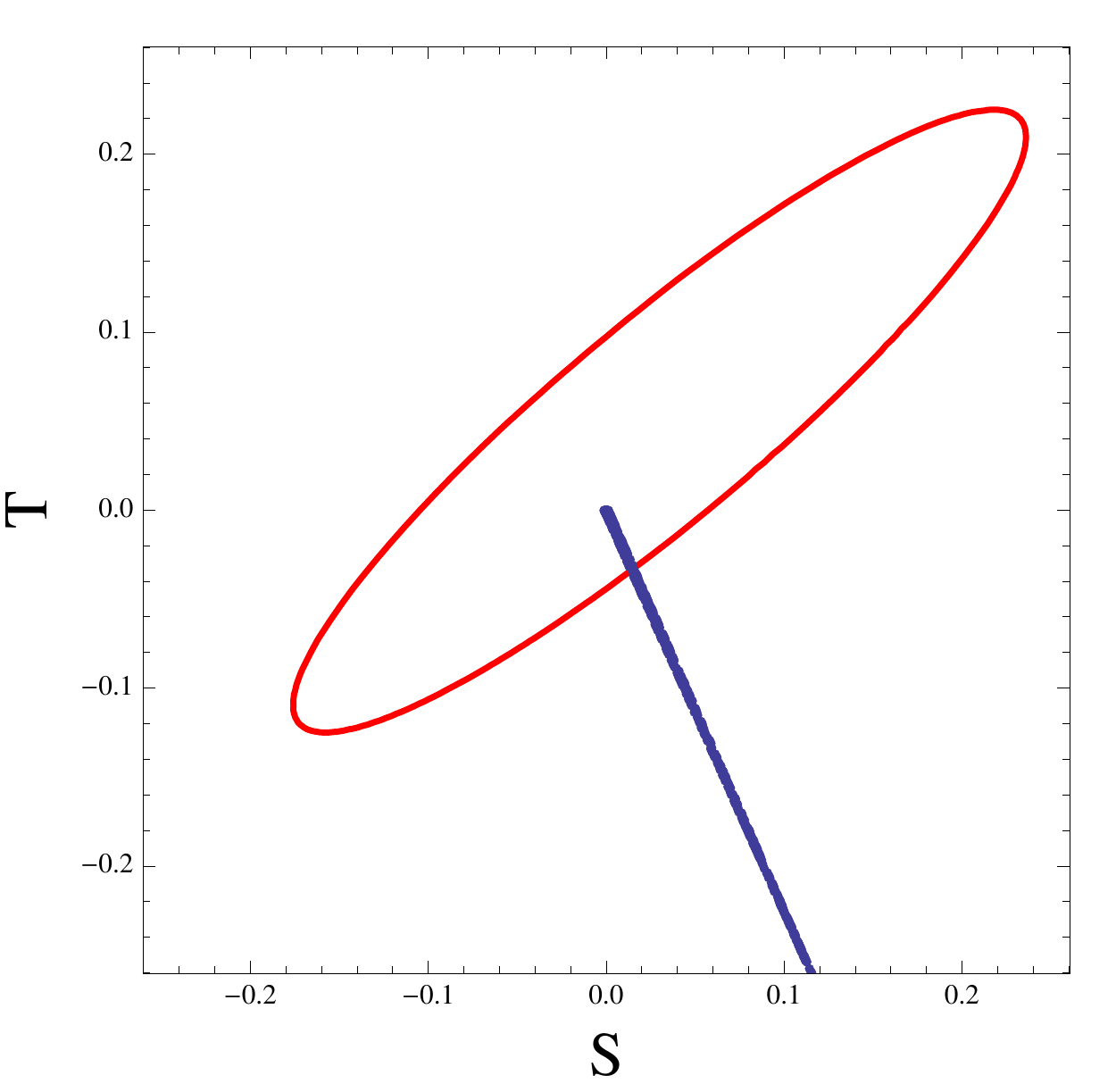}
\hfill
\includegraphics[width=0.46\textwidth]{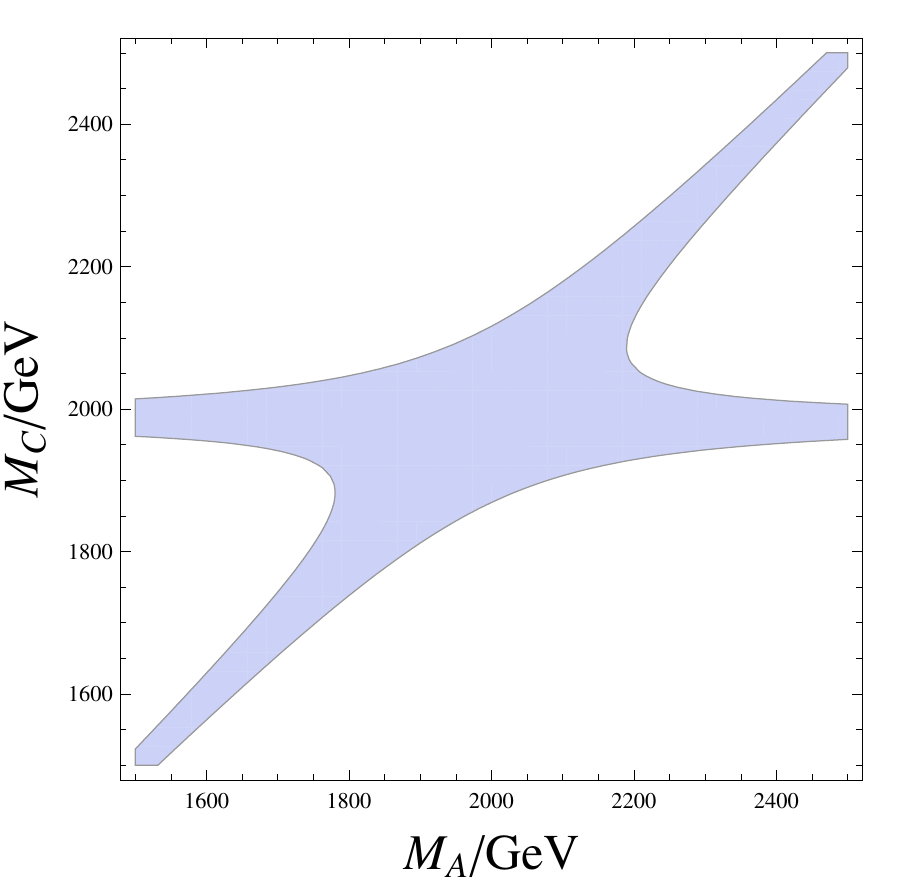}
\caption{  Left panel: Predictions of heavy state in the $S-T$ plane by setting $M_C=M_A$ and $M_H=2~{\rm TeV}$; Right panel: Constraints on the masses of the charged and CP-odd neutral states from oblique parameters by setting $m_H=2~{\rm TeV}$ and $\sin \alpha \sim 0.1$. }
\label{fig:stu}
\end{figure}

The most recent electroweak fit (by setting  $m_{h,ref} =126~{\rm GeV}$ and $m_{t,ref}=173~{\rm GeV}$) to the oblique parameters performed by the Gfitter group~\cite{Baak:2012kk} yields 
\begin{eqnarray}
S\equiv \Delta S^0 \pm \sigma_S=0.03\pm0.10\; , \hspace{1cm} T\equiv \Delta T^0 \pm \sigma_T =0.05\pm0.12 \; .
\end{eqnarray}
The $\Delta \chi^2 $ can be written as 
\begin{eqnarray}
\Delta \chi^2 =\sum_{ij}^2 (\Delta {\cal O}_i - \Delta {\cal O }_i^0 )( \sigma^2 )_{ij}^{-1} (\Delta {\cal O}_j  - \Delta {\cal O}_j^0) 
\end{eqnarray}
where ${\cal O}_1=S$  and ${\cal O}_2=T$; $\sigma^2_{ij} =\sigma_i \rho_{ij} \sigma_j$ with $\rho_{11}=\rho_{22}=1$ and $\rho_{12}=0.891$.  

As can be seen from eqs. (\ref{scal}) and (\ref{tcal}), there are four free parameters contributing to the oblique parameters, $m_A$, $m_C$, $m_H$ and $\alpha$. To preform electroweak fit, we set $M_C=M_A\equiv M$, which can be easily achieved by setting $\lambda_3=\lambda_4$, and $m_H=2~{\rm TeV}$, so that only two free parameters left.  Blue points In the left panel of FIG.\ref{fig:stu} show the contribution to the $\Delta S$ and $\Delta T$ by setting $M$ and $\sin \alpha$ random parameters varying in the range $ (1.8,~2.3)~{\rm TeV}$   and $(0,~1)$ respectively. The contour in the same plot shows the allowed region in the $S-T$ plane in the $95\%$ C.L.  A direct numerical calculation shows that $|\sin \alpha| \leq 0.3$. In the right panel of FIG. \ref{fig:stu} we show the region that are allowed by the oblique observations in the $M_C-M_A$ plane by setting $\sin\alpha =0.1$ and $M_H=2~{\rm TeV}$.  To summarize, electroweak precision measurements put stronger constraint on the $\alpha$ even for the nearly degenerate heavy states.  

\begin{figure}[t!]
\centering
\includegraphics[width=0.46\textwidth]{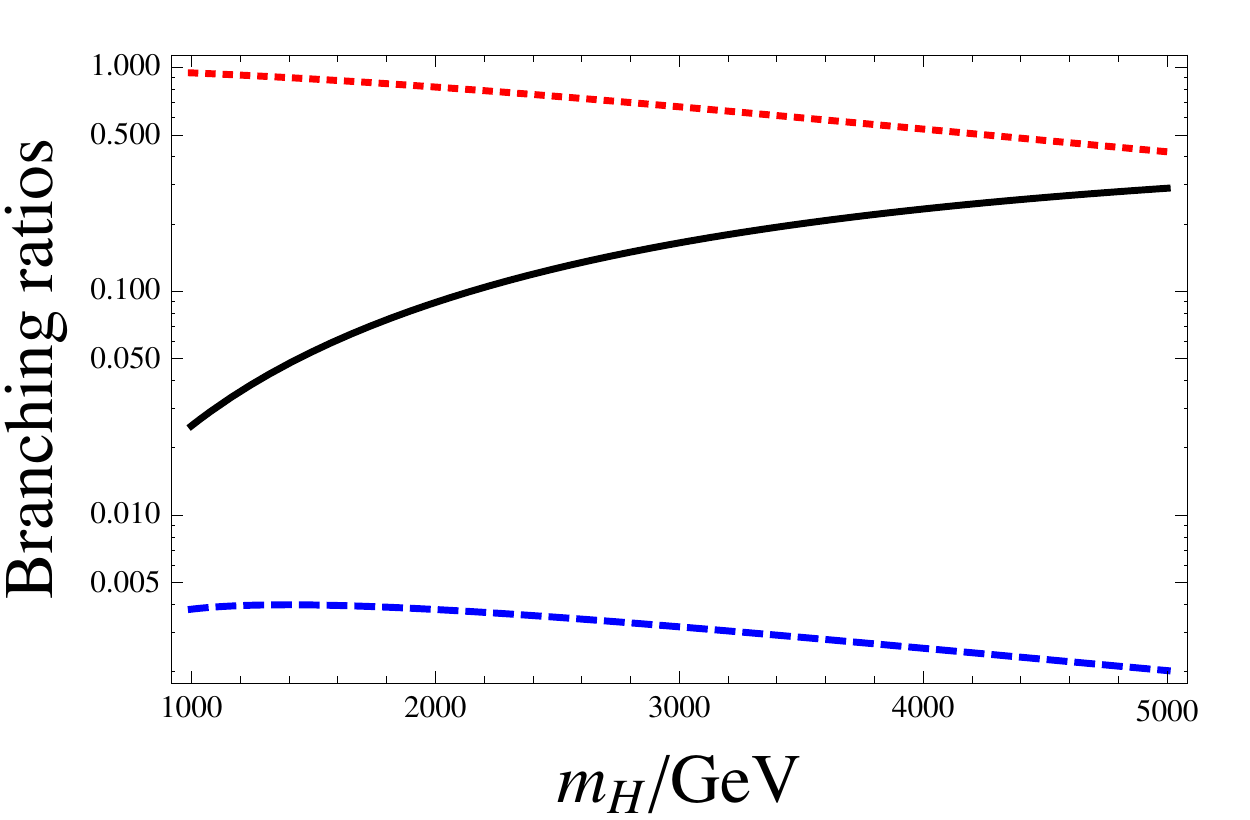}
\hfill
\includegraphics[width=0.46\textwidth]{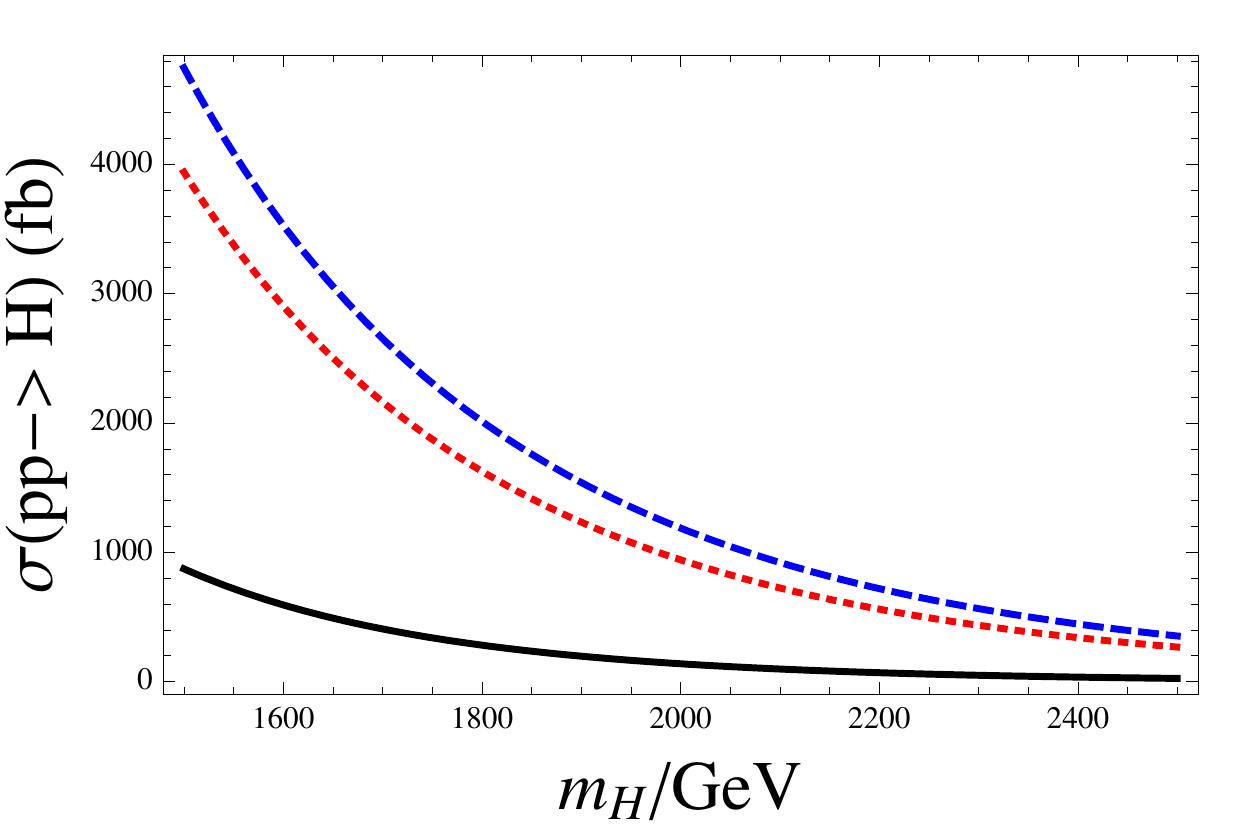}
\caption{  Left panel: Branching ratios of $H$ as the function of $m_H$ by setting $s\sim 0.05$ and $\xi=0.5$; Right panel: Production cross section of the heavy CP-even Higgs boson at the LHC by setting $\xi=0.5$, with solid, dotted and dashed lines correspond to $\sqrt{s}=8,~13,~14~{\rm TeV}$ respectively. }
\label{fig:BR&PRO}
\end{figure}

\section{diboson excesses}

Heavy scalar states in our model can be produced at the LHC through its Yukawa interaction with the first generation quarks as was shown in eq. (\ref {yukawah}) and can decay into diboson final states from the mixing with the SM-like Higgs boson.  The main decay channels of $H$ are $\bar u u$, $\bar t t$, $W^+W^-$ and $ZZ$. The decay rates can be written as 
\begin{eqnarray}
\Gamma_{u\bar u} ~&=& {n_C \xi^2 m_H \over 8\pi} \label{dr1}\\
\Gamma_{t\bar t}~~ &=& { s^2 n_C m_t^2 (m_H^2 -m_t^2 )^{3/2} \over 8\pi m_H^2 v^2 } \\
\Gamma_{VV} &=& { s^2 m_V^4 \sqrt{m_H^2 -m_V^2 } \over 4 \pi m_H^2 v^2 } \left( 3 - {m_H^2 \over m_V^2 } + {m_H^4 \over 4 m_V^4}\right)
\end{eqnarray}
where $n_C=3$, being the color index; $V=W,~Z$ respectively.  We show in the left panel of FIG. \ref{fig:BR&PRO} the branching ratios of $H$ by setting $s=0.05$ and $\xi=0.5$, where the solid, dotted and dashed lines correspond to the branching ratios of  $WW/ZZ$, $\bar u u$ and $\bar t t$ respectively.  We plot in the right panel of FIG. \ref{fig:BR&PRO}  the production cross section of $H$ at the LHC. The solid, dotted and dashed lines correspond to $\sqrt{s}=8~{\rm TeV},~13~{\rm TeV}$ and $14~{\rm TeV}$, respectively.

 \begin{figure}[t!]
\centering
\includegraphics[width=0.46\textwidth]{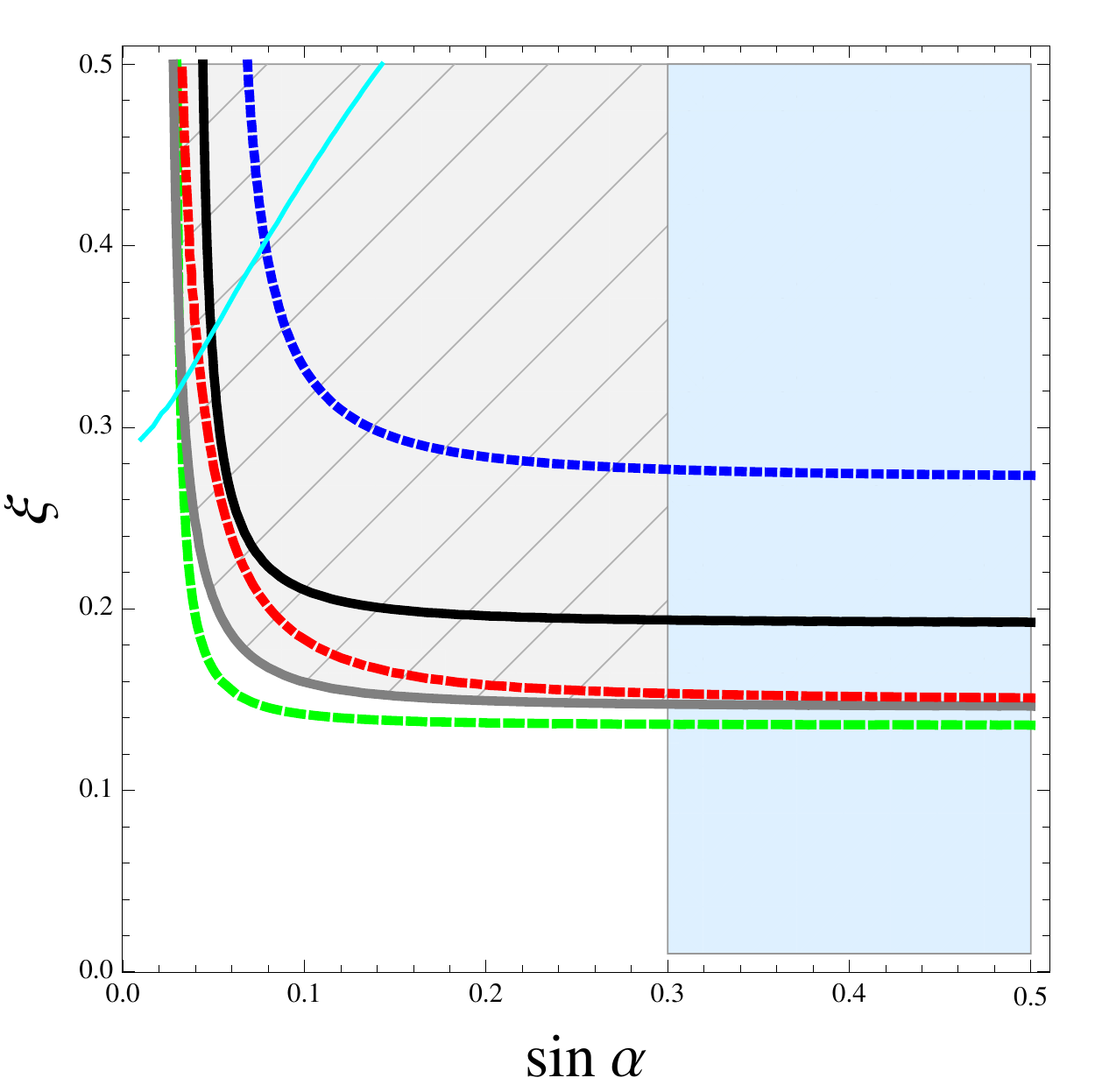}
\caption{ Contour plot of $\sigma(pp\to H \to WW)$ in the $\sin\alpha-\xi$ plane. The blue dotted, black solid and green dashed lines correspond to $\sigma (pp\to H \to WW) =20, ~10, ~5 ~{\rm fb^{}}$ respectively. The region below the gray solid line satisfies $ \sigma(pp\to A\to hZ) <7~{\rm fb}$. The region below the red dot-dashed line satisfies $ \sigma(pp\to C \to hW ) <7~{\rm fb}$. The region below the cyan solid line has $\sigma (pp\to R\to jj ) <100~{\rm fb}$.  }
\label{fig:CROSSS}
\end{figure}

We show in FIG. \ref{fig:CROSSS} the contours of $\sigma(pp\to H\to WW)$ in the $\sin \alpha -\xi$ plane. The dashed, solid and dotted lines correspond to  $\sigma(pp\to H\to WW)=5,~10,~20~{\rm fb^{}}$ respectively.  One can get similar numerical results for the $(pp\to H\to ZZ)$ process. The ATLAS reported number of excesses is about $8\sim 9$ events near the $2~{\rm TeV}$ peak.  Given a luminosity of $20.3 ~{\rm fb^{-1}}$, one has $ \sigma(pp\to H\to WW)\approx 5\sim 6 ~{\rm fb^{}}$ for a $13\%$ \cite{Aad:2015owa} selection efficiency of the event topology and boson-tagging requirements.  Although large enough cross section can be produced at the LHC, the model is constrained by other LHC experimental results. We will discuss these constraints one-by-one as follows: 
\begin{itemize}
\item  The CMS collaboration~\cite{Khachatryan:2015bma} has reported an upper bound for the $\sigma(pp\to R\to W^+ h)$, where $R$ is a new resonance. It gives $\sigma (pp\to R \to W^+ h) \leq 7 ~{\rm fb^{}}$. The resonance can be the charged component of the heavy scalar doublet in our model. Its decay rate can be written as
\begin{eqnarray}
\Gamma_{C\to Wh}&=&{g^2 s^2 \over 64 \pi m_W^2 m_C^3 } \lambda^{3/2} (m_C^2, m_h^2, m_W^2 ) \; , 
\end{eqnarray}
where $\lambda(x,~y,~z)=x^2 + y^2 +z^2 -2xy-2xz-2yz$ and $g$ is the $SU(2)$ gauge coupling.  FIG. \ref{fig:CROSSS}  the numerical results by setting $m_C=2.2~{\rm TeV}$,  where the region below the red dot-dashed line satisfy this constraint.

\item The CP-odd component of the heavy scalar doublet can be the mediator of the process $pp\to R \to Zh $, which was also measured the CMS collaboration. One has $\sigma(pp\to A \to Zh)<7~{\rm fb^{}}$  The decay rate of $A\to Z h $ can be written as
\begin{eqnarray}
\Gamma_{A\to Z h} = {g^2 s^2 \over 64 \pi c_W^2 m_Z^2 m_A^3} \lambda^{3/2} (m_A^2, m_Z^2, m_h^2 )
\end{eqnarray}
where $c_W =\cos \theta_W$ with $\theta_W$ the weak mixing angle.  $A$ can also decay into dijet final states with the decay rate the same as eq. (\ref{dr1}).  We show in FIG. \ref{fig:CROSSS}  the numerical results,  where the region to the top-right of  the gray solid line are excluded by this constraint. 

\item  Both ATLAS and CMS has searched for resonances decaying into dijets.  We use $\sigma(pp\to R\to jj)\leq 100~{\rm fb^{}}$ with the acceptance ${\cal A }\sim 0.6$.   Both the CP-even and the CP-odd heavy scalars as well as the charged scalar  in our model mainly decay into dijet via the Yukawa interaction.  We show in FIG. \ref{fig:CROSSS} the region (to the bottom-right corner of the cyan solid line) allowed by this constraint. 
\end{itemize}
Since the decay rate of $H\to t\bar t$ is tiny, there is almost no constraint on the model from $t\bar t$ resonance searches.  As can be seen from FIG. \ref{fig:CROSSS}, $\sigma(pp\to H \to WW)$ should be less than $6\sim 7$ fb. One has $\sigma(pp\to H \to WW/ZZ) \sim 5~{\rm fb} $ for $\xi\sim 0.15$ and $ \alpha \sim  0.06$, which is consistent with the constraints of colliders and electroweak precision measurements. No direct excess in the $WZ$ channel comes out of our model. But the the ATLAS observed excess in the $WZ$ channel can be interpreted as the misidentification of the $W/Z$-tagged jet owing to uncertainties of the tagging  slections.

\section{Summary}

We investigated the prospects of the stealth doublet model as a possible explanation to the diboson excesses observed by the ATLAS collaboration. The mass of heavy Higgs boson was fixed at near $2~{\rm TeV}$ in our study.  We showed that excesses in the $WW$ and $ZZ$ channels can be interpreted as the decay of the heavy CP-even Higgs boson $H$, which can be produced at the LHC via its Yukawa interaction with the first generation quarks.   One needs the Yukawa coupling  $\xi\sim0.15$ and the mixing angle between two CP-even Higgs bosons  $\alpha\sim0.06$, which is consistent with precision measurements, so as to has a $5~{\rm fb}$ production cross section at the LHC.  Constraints on the model from the exclusion limits in $Wh$ and $Zh$ channels given by CMS collaboration and dijet searches was also studied,  which showed the limited  parameter space (in FIG. \ref{fig:CROSSS}) that can be accommodated with the interpretation of the ATLAS diboson excesses in the same model. We expect the running of the 13 TeV LHC to tell us the detail about the diboson  excesses and show us more clear hints of new physics behind this phenomena.

\begin{acknowledgments}
The author thanks to Huaike Guo and Peter Winslow for very helpful discussions.
This work was supported in part by DOE Grant DE-SC0011095

\end{acknowledgments}


\begin{thebibliography}{99}

\bibitem{Aad:2015owa} 
  G.~Aad {\it et al.}  [ATLAS Collaboration],
  arXiv:1506.00962 [hep-ex].

\bibitem{Khachatryan:2014hpa} 
  V.~Khachatryan {\it et al.} [CMS Collaboration],
  JHEP {\bf 1408}, 173 (2014)
  [arXiv:1405.1994 [hep-ex]].
  
  
\bibitem{Khachatryan:2014gha} 
  V.~Khachatryan {\it et al.} [CMS Collaboration],
  JHEP {\bf 1408}, 174 (2014)
  [arXiv:1405.3447 [hep-ex]].

 
 
 
  
\bibitem{Hisano:2015gna} 
  J.~Hisano, N.~Nagata and Y.~Omura,
  arXiv:1506.03931 [hep-ph].
  








\bibitem{Fukano:2015hga} 
  H.~S.~Fukano, M.~Kurachi, S.~Matsuzaki, K.~Terashi and K.~Yamawaki,
  arXiv:1506.03751 [hep-ph].
  
\bibitem{Franzosi:2015zra} 
  D.~B.~Franzosi, M.~T.~Frandsen and F.~Sannino,
  arXiv:1506.04392 [hep-ph].
  
\bibitem{Cheung:2015nha} 
  K.~Cheung, W.~Y.~Keung, P.~Y.~Tseng and T.~C.~Yuan,
  arXiv:1506.06064 [hep-ph].
  
\bibitem{Dobrescu:2015qna} 
  B.~A.~Dobrescu and Z.~Liu,
  arXiv:1506.06736 [hep-ph].
  
\bibitem{Aguilar-Saavedra:2015rna} 
  J.~A.~Aguilar-Saavedra,
  arXiv:1506.06739 [hep-ph].
  
\bibitem{Gao:2015irw} 
  Y.~Gao, T.~Ghosh, K.~Sinha and J.~H.~Yu,
  arXiv:1506.07511 [hep-ph].
  
\bibitem{Thamm:2015csa} 
  A.~Thamm, R.~Torre and A.~Wulzer,
  arXiv:1506.08688 [hep-ph].
  
\bibitem{Brehmer:2015cia} 
  J.~Brehmer, J.~Hewett, J.~Kopp, T.~Rizzo and J.~Tattersall,
  arXiv:1507.00013 [hep-ph].
  
\bibitem{Cao:2015lia} 
  Q.~H.~Cao, B.~Yan and D.~M.~Zhang,
  arXiv:1507.00268 [hep-ph].
  
\bibitem{Cacciapaglia:2015eea} 
  G.~Cacciapaglia and M.~T.~Frandsen,
  arXiv:1507.00900 [hep-ph].
  
\bibitem{Abe:2015jra} 
  T.~Abe, R.~Nagai, S.~Okawa and M.~Tanabashi,
  arXiv:1507.01185 [hep-ph].
  
\bibitem{Allanach:2015hba} 
  B.~C.~Allanach, B.~Gripaios and D.~Sutherland,
  arXiv:1507.01638 [hep-ph].
  
\bibitem{Abe:2015uaa} 
  T.~Abe, T.~Kitahara and M.~M.~Nojiri,
  arXiv:1507.01681 [hep-ph].
  
  
\bibitem{Carmona:2015xaa} 
  A.~Carmona, A.~Delgado, M.~Quiros and J.~Santiago,
  arXiv:1507.01914 [hep-ph].
  
  
\bibitem{Dobrescu:2015yba} 
  B.~A.~Dobrescu and Z.~Liu,
  arXiv:1507.01923 [hep-ph].
  







\bibitem{Chiang:2015lqa} 
  C.~W.~Chiang, H.~Fukuda, K.~Harigaya, M.~Ibe and T.~T.~Yanagida,
  arXiv:1507.02483 [hep-ph].
  
  
\bibitem{Cacciapaglia:2015nga} 
  G.~Cacciapaglia, A.~Deandrea and M.~Hashimoto,
  arXiv:1507.03098 [hep-ph].
  
\bibitem{Sanz:2015zha} 
  V.~Sanz,
  arXiv:1507.03553 [hep-ph].
  
\bibitem{Chen:2015xql} 
  C.~H.~Chen and T.~Nomura,
  arXiv:1507.04431 [hep-ph].
  
  
  
\bibitem{Enberg:2013jba} 
  R.~Enberg, J.~Rathsman and G.~Wouda,
  Phys.\ Rev.\ D {\bf 91}, no. 9, 095002 (2015)
  [arXiv:1311.4367 [hep-ph]].
  
\bibitem{Enberg:2013ara} 
  R.~Enberg, J.~Rathsman and G.~Wouda,
  JHEP {\bf 1308}, 079 (2013)
  [JHEP {\bf 1501}, 087 (2015)]
  [arXiv:1304.1714 [hep-ph]].
  
\bibitem{Profumo:2014opa} 
  S.~Profumo, M.~J.~Ramsey-Musolf, C.~L.~Wainwright and P.~Winslow,
  Phys.\ Rev.\ D {\bf 91}, no. 3, 035018 (2015)
  [arXiv:1407.5342 [hep-ph]].
  
  
\bibitem{Peskin:1990zt} 
  M.~E.~Peskin and T.~Takeuchi,
  Phys.\ Rev.\ Lett.\  {\bf 65}, 964 (1990).
  
\bibitem{Peskin:1991sw} 
  M.~E.~Peskin and T.~Takeuchi,
  Phys.\ Rev.\ D {\bf 46}, 381 (1992).
  
  
\bibitem{Haber:2010bw} 
  H.~E.~Haber and D.~O'Neil,
  Phys.\ Rev.\ D {\bf 83}, 055017 (2011)
  [arXiv:1011.6188 [hep-ph]].
  
  
\bibitem{Baak:2012kk} 
  M.~Baak, M.~Goebel, J.~Haller, A.~Hoecker, D.~Kennedy, R.~Kogler, K.~Moenig and M.~Schott {\it et al.},
  Eur.\ Phys.\ J.\ C {\bf 72}, 2205 (2012)
  [arXiv:1209.2716 [hep-ph]].
  
  
\bibitem{Khachatryan:2015bma} 
  V.~Khachatryan {\it et al.} [CMS Collaboration],
  arXiv:1506.01443 [hep-ex].
  
\end{thebibliography}
\end{document}